# Investigations of the Core Structure of Magnetic Vortices in Type-II Superconductors by μSR

Jeff E. Sonier
Department of Physics, Simon Fraser University, Burnaby, British Columbia V5A 1S6, Canada

**Abstract**

Muon spin rotation (μSR) has emerged as the leading experimental probe of the effective size of magnetic vortices in type-II superconductors. μSR data on several different classes of type-II superconductors shows that the inner structure of a vortex can depend quite strongly on temperature and the strength of the external magnetic field. In this paper it is shown that these behaviours are related to the quasiparticle excitation spectrum both inside and outside of the vortex cores. Here we establish that the vortex-core size determined by μSR is particularly sensitive to the nature of the superconducting energy gap(s) and the symmetry of the Fermi surface. A survey of results for different superconductors arrives at the conclusion that the large vortex-core size observed by μSR in $YBa_2Cu_3O_{7-\delta}$ is due to CuO chain superconductivity.

## 1. Introduction

Alexei Abrikosov shared the 2003 Nobel Prize in physics for his astounding prediction 46 years ago, that some materials could retain superconductivity in strong magnetic fields by allowing the external field to penetrate the sample as a periodic arrangement of quantized magnetic flux lines, *i.e.* a *vortex lattice* (VL) [1]. Since Abrikosov's initial theoretical description of the vortex state, there has been tremendous effort directed toward understanding the behaviour of magnetic flux lines (vortices) in type-II superconductors.

The basis of Abrikosov's model of the vortex state is the phenomenological Ginzburg-Landau (GL) theory, which simplifies the theoretical treatment of a spatially varying order parameter $\Delta(r)$ in a type-II superconductor [2]. Within the framework of the GL theory the structure of an Abrikosov vortex is characterized by two fundamental length scales: the *magnetic penetration depth* $\lambda$, and the *coherence length* $\xi$, where $\lambda > \xi$. The coherence length $\xi$ is approximately the distance over which the pair potential (order parameter) $\Delta(r)$ rises from zero at the center of a vortex ($r = 0$) to its asymptotic value $\Delta_0$ outside the vortex core. Its value can be determined experimentally from measurements of the upper critical field, $H_{c2} = \Phi_0/2\pi\xi^2$. The density of the supercurrent $j(r)$ also rises steeply in the core region. Beyond its maximum, the absolute value of $j(r)$ decays over the length scale $\lambda$. Strictly speaking, GL theory is valid only near the superconducting transition temperature $T_c$, where $\Delta(r)$ varies slowly in space. In this regime, the GL equations are derivable from the microscopic Bardeen, Cooper, Schrieffer (BCS) theory [3] that explains superconductivity in simple elements and alloys [4]. On the other hand, well below $T_c$ where the spatial variation of $\Delta(r)$ is rapid and GL theory breaks down, a proper theoretical description of the vortex state necessitates solutions of the microscopic theory. Even so, the major effort that is required to solve the microscopic equations for a spatially varying order parameter has resulted in widespread reliance on the GL theory to describe VLs at low temperatures. However, this simplification of the problem does not account for the quasiparticle excitation spectrum of the vortex cores.

In recent years, there have been great strides made toward understanding *vortex-vortex interactions* in the framework of the microscopic theory. The microscopic equations couple the spatial variation of the internal magnetic field $B(\mathbf{r})$ to the electronic spectrum of the vortex cores. The µSR technique can probe $B(\mathbf{r})$ in the bulk of a superconductor in the vortex state. Consequently, µSR measurements can determine the vortex-core size and hence provide an experimental means of confirming the role of quasiparticles in the vortex state.

## 2. Vortex-Core Size

Caroli, de Gennes and Matricon first formally treated the problem of an isolated vortex in a clean conventional *s*-wave superconductor within the framework of the microscopic theory [5]. Using the Bogoliubov-de Gennes (BdG) equations they showed that quasiparticles whose energy $E$ is less than the bulk energy gap $\Delta_0$ form discrete bound states in the vortex core. The bound states arise from repeated Andreev scattering of the quasiparticles from the pair potential $\Delta(r)$ in the core region [6]. The bound states have wavefunctions that are exponentially localized near $r = 0$, such that the higher-energy states extend out to larger radii. Gygi and Schlüter showed in the framework of the BdG theory that quasiparticles with energies $E > \Delta_0$ also contribute to the spatial structure of a vortex line [7]. These quasiparticles scatter from the vortex and dominate the behaviour of $\Delta(r)$ far from the vortex center.

A direct consequence of the electronic spectrum of the vortex core is that its inner structure depends on temperature and magnetic field. At low temperature and/or large magnetic field, the pair potential can rise rapidly at the vortex center. In these cases the commonly used analytic relation $\Delta(r) = \Delta_0 \tanh(r/\xi_0)$, where $\xi_0 \sim 1/\Delta_0$ is the BCS coherence length, does not accurately describe the spatial dependence of the order parameter at the vortex site. Instead, numerical solutions of the microscopic theory are required. Theoretical works in the framework of the BdG or Eilenberger theories define the vortex-core size either from the spatial variation of the pair potential $\Delta(r)$ or the supercurrent density $j(r)$ near the vortex center (see Fig. 1). From the slope of $\Delta(r)$ at the vortex center, one may define the core radius as

$$\xi_1 = \Delta_0 / \lim_{r \to 0} \frac{\Delta(r)}{r} \quad . \tag{2.1}$$

Alternatively, one can define the core radius to be the distance $r = r_0$ from the core center at which the absolute value of the supercurrent density $j(r)$ reaches its maximum value.

Since $\Delta(r)$ and $j(r)$ are related through a self-consistent iterative procedure, $\xi_1$ and $r_0$ exhibit qualitatively similar behaviours as a function of temperature and magnetic field. It is important to note that the radii $\xi_1$ and $r_0$ are generally much smaller than both the BCS coherence length $\xi_0$ and the GL coherence length $\xi$ determined from $H_{c2}$, i.e. $\xi_{Hc2} = (\Phi_0/2\pi H_{c2})^{1/2}$.

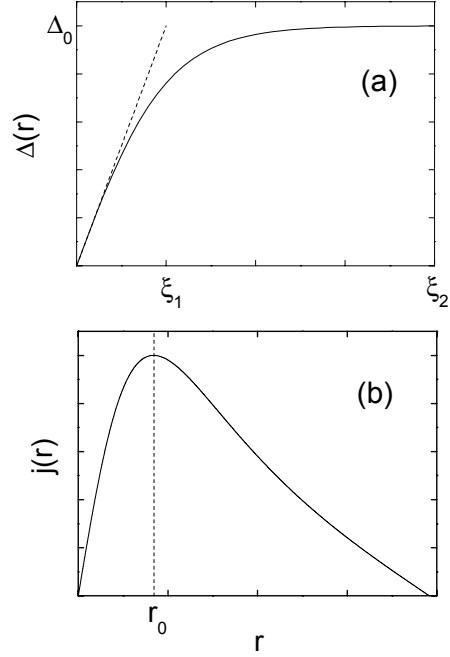

**Figure 1** Spatial variation of (a) the pair potential $\Delta(r)$ and (b) the absolute value of the supercurrent density $j(r)$ for an isolated vortex, where $r$ is the radial distance from the vortex center. The length scales $\xi_1$, $\xi_2$ and $r_0$ are defined in the text.

Gygi and Schlüter [7], showed that $\Delta(r)$ at a vortex site is governed by $\xi_1$ and a second length scale $\xi_2$, such that $\Delta(r) = \Delta_0\tanh(r/\xi_2)$ and $\xi_2 \leq \xi_0$. There are two important points to emphasize here. The first is that $\xi_1$ and $\xi_2$ are parameters that provide an analytical means of characterizing what is generally a numerical function. In other words, they are not fundamental length scales that appear in the microscopic theory. The second point to make is that the length scales $\xi_1$, $\xi_2$ and $r_0$ can vary quite strongly with temperature and/or magnetic field, even when the BCS coherence length $\xi_0$ does not change appreciably. Hence, variations of the "vortex-core size" defined by $\xi_1$ or $r_0$ imply neither a change in the size of the Cooper pairs nor the upper critical field $H_{c2}$.

## 2.1 Temperature Dependence

Kramer and Pesch showed that thermal depopulation of the higher-energy bound core states leads to a shrinking of the vortex core radius with decreasing temperature $T$ [8,9]. The reduction in core size occurs because the lower-energy bound state wave functions do not extend as far out from the vortex center as the high-energy ones. The decrease of $\xi_1$ ceases when only the lowest bound core state is populated. These theoretical ideas have been confirmed in calculations by others [7,10,11]. Nevertheless, it is important to note that predictions for the dependence of the core size on temperature exist only for the case of an isolated or weakly interacting vortex. Thus, only qualitative agreement with µSR results is expected.

## 2.2 Magnetic Field Dependence

With increased applied magnetic field $H$, the density of vortices increases. Calculations of $r_0$ from the analytical approximations of the field profile $B(r)$ in the London [12] and GL [13] models show that $r_0 \geq \xi$, and that $r_0$ decreases with increasing $H$. The latter arises because of the superposition of the supercurrent density profiles $j(r)$ associated with individual vortices. As shown in Fig. 2, the supercurrent density profiles of nearest-neighbour vortices have opposite signs. The sum of the individual $j(r)$ profiles shifts the maximum in the absolute value of $j(r)$ around a vortex closer to the vortex center. Increasing the magnetic field brings the vortices closer together, which increases the overlap of the $j(r)$ profiles in Fig. 2 and hence reduces $r_0$. We will refer to this as the "vortex-lattice squeezing effect".

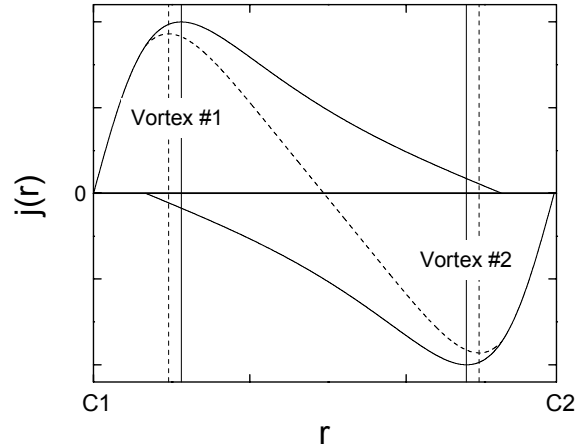

**Figure 2** Superposition of the supercurrent density profiles $j(r)$ of nearest-neighbour vortices. The solid curves represent the $j(r)$ profiles of two individual vortices, where C1 and C2 denote the vortex centers, and the solid vertical lines indicate the maximum in the absolute value of $j(r)$. The dashed curve is the resultant $j(r)$ and the dashed vertical lines indicate the corresponding new positions of the maximums.

There is an additional contribution to the field dependence of $r_0$, which is a consequence of the quasiparticle excitation spectrum of a vortex. At low fields, where the vortices are far apart, the bound core states of nearest-neighbour vortices weakly overlap. In this regime the vortex-vortex interactions are primarily governed by isotropic magnetic repulsion. The reduction in the intervortex spacing $L \sim 1/H^{1/2}$ that occurs with increasing $H$ leads to a stronger overlap of the bound state wave functions of neighbouring vortices, forming energy bands that enable the intervortex transfer of quasiparticles [14-18]. This delocalization of quasiparticle core states modifies the spatial variation of $\Delta(r)$, and hence the size of the vortex cores defined by $\xi_1$ and $\xi_2$. Golubov and Hartmann [19] first showed in the dirty limit of the microscopic theory that the vortex-core size in a conventional *s*-wave superconductor shrinks with increasing *H*. Subsequently this behaviour was verified in the clean-limit [17,18].

## 2.3 Gap and Fermi Surface Anisotropy

In real materials the vortex structure is also strongly influenced by both the anisotropy of the energy gap and the anisotropy of the Fermi surface. The coupling of the VL to the underlying crystal lattice affects both the symmetry of the vortex cores and their orientation. Kogan *et al.* [20] have modified the phenomenological London model to account for the nonlocality of the relation between the supercurrent density **j**(*r*) and the vector potential **A** in the region around the vortex core. The nonlocal London model couples vortex structure to the anisotropy of the Fermi surface, and has been successful in explaining field-induced symmetry changes of the VL in strong type-II superconductors. In general, VLs are hexagonal at low fields, where the intervortex spacing is large and the vortices interact predominantly by isotropic magnetic repulsion. With increasing field, the increased overlap of neighbouring vortices eventually causes the VL to adopt the symmetry of the individual vortices. More recently, Nakai *et al.* [21] have developed a generalized model for VL transformations in the framework of the quasiclassical Eilenberger theory, which includes the effect of the superconducting energy gap. The symmetry of the VL in this model reflects a competition between the anisotropies of the energy gap and the Fermi surface.

## 3. Muon Spin Rotation (μSR)

Almost everything known experimentally about the behaviour of the vortex-core size in type-II superconductors has come from μSR studies. Surface-sensitive techniques, such as scanning tunneling spectroscopy (STS) and scanning SQUID microscopy, are less accurate because the vortex lines expand considerably at the sample surface [22,23]. In a μSR experiment the core size is determined from the measured internal magnetic field distribution *n*(*B*) [24]. In principle, one can also measure *n*(*B*) by conventional nuclear magnetic resonance (NMR). However, unlike μSR, NMR is not a pure magnetic probe (*i.e.* it is simultaneously sensitive to electrostatic fields). Furthermore, NMR requires an RF field, which for the case of a large single crystal is effectively screened from the bulk by the supercurrents flowing around the perimeter of the sample.

A major consideration in the μSR approach is that some modeling of *n*(*B*) is required. The measured time evolution of the muon spin polarization *P*(*t*) is fit to a theoretical function

$$P(t) = G(t)\int_0^\infty n(B)\cos(\gamma_\mu Bt + \theta)dB \, , \qquad (3.1)$$

where $B$ is the local field sensed by the muon, $\gamma_\mu$ is the muon gryomagnetic ratio, $\theta$ is the initial phase of the muon spin polarization vector $P(0)$, and $G(t)$ is a depolarization function appropriately chosen to account for additional sources of field inhomogeneity, such as VL disorder and randomly oriented magnetic moments. The internal magnetic field distribution is the probability of finding a field $B$ at an arbitrary position $r$, such that [12]

$$n(B') = \langle \delta[B' - B(\mathbf{r})] \rangle \, , \tag{3.2}$$

where $\langle \ldots \rangle$ denotes spatial averaging and $\delta$ is the one-dimensional delta function. The spatial variation of the magnetic field due to a periodic arrangement of vortices is

$$B(\mathbf{r}) = \sum_{\mathbf{G}} B_{\mathbf{G}} e^{-i\mathbf{G}\cdot\mathbf{r}} \, . \tag{3.3}$$

The sum is over the reciprocal lattice vectors $\mathbf{G}$ of the unit cell. The choice of a model for the Fourier components $B_{\mathbf{G}}$ depends on the problem at hand. For an extreme type-II superconductor a good approximation is the analytical GL model [13]

$$B_{\mathbf{K}} = B_0 (1 - b^4) \frac{u K_1(u)}{\lambda^2 |\mathbf{G}|^2} \, , \tag{3.4}$$

where $B_0$ is the average internal field, $b = B/H_{c2}$, $K_1(u)$ is a modified Bessel function, and $u^2 = 2\xi^2 |\mathbf{G}|^2 (1 + b^4)[1 - 2b(1 - b)^2]$. We note that Eq. (3.4) accounts for the *finite size* of the vortex cores. This is required to accurately determine the magnetic penetration depth $\lambda$. For a field applied parallel to the $\hat{c}$-axis of a crystal we denote $\lambda \equiv \lambda_{ab}$ and $\xi \equiv \xi_{ab}$. Strictly speaking, $\lambda_{ab}$ and $\xi_{ab}$ are fitting parameters whose temperature and magnetic field dependence characterize changes of $B(\mathbf{r})$. They are true measures of these physical length scales only in situations where one is confident that the model used for $B(\mathbf{r})$ is precise. For example, in highly anisotropic superconductors, nonlinear and nonlocal effects may strongly influence $B(\mathbf{r})$ [25]. Unless $B(\mathbf{r})$ accounts for these effects, the measured $\lambda_{ab}(T, H)$ will not accurately reflect the intrinsic behaviour of the superfluid density.

Likewise, the measured $\xi_{ab}(T, H)$ does not reflect the behaviour of $H_{c2}$ nor the size of the Cooper pairs. In the microscopic theory, changes in $\Delta(r)$ alter $B(r)$ over the same spatial region [18]. Consequently, we expect $\xi_{ab}$, which is related to $B(\mathbf{r})$ near the vortex core region, to be sensitive to changes in $\xi_1$ and/or $\xi_2$. Since the absolute value of $\xi_{ab}$ depends somewhat on the model for $B(\mathbf{r})$, it has become customary to determine the vortex-core size from the absolute value of the supercurrent density profile calculated using Maxwell's relation: $j(r) = |\nabla \times \mathbf{B}(r)|$ (Note: $r$ is the position along the straight line connecting nearest-neighbour vortices and $\mathbf{B}(r)$ is parallel to the applied field direction). Reliable values of $r_0$ (see Fig. 1(b) for definition) can be obtained in this way independent of the validity of $B(\mathbf{r})$, provided a good fit of the data is achieved. In general, the field dependence of $r_0$ will consist of contributions from the vortex-lattice squeezing effect and the intervortex transfer of quasiparticles. On the other hand, it will be shown that the fitting parameter $\xi_{ab}$ is most sensitive to the delocalization of core states and changes in the superconducting energy gap. Table 1 contains a summary of the different length scales introduced here in the description of the vortex-core size.

**Table 1** Theoretical and experimental definitions of the length scales used to describe vortex-core size.

|  | Length Scale | Definition |
|---|---|---|
| **Theoretical:** | $\xi_0$ | BCS coherence length |
|  | $\xi$ | Ginzburg-Landau (GL) coherence length |
|  | $\xi_1$ | $\xi_1 = \Delta_0 / \lim_{r \to 0} \frac{\Delta(r)}{r}$ |
|  | $\xi_2$ | $\Delta(r) = \Delta_0 \tanh\left(\frac{r}{\xi_2}\right)$ |
|  | $\xi_{Hc2}$ | $H_{c2} = \frac{\Phi_0}{2\pi \xi_{Hc2}^2}$ (from GL theory) |
|  | $r_0$ | Position of the maximum value of the supercurrent density $j(r)$ relative to the vortex center, measured along the straight line connecting nearest-neighbour vortices. [Note: behaviour of $r_0$ as a function of $T$ and $H$ is qualitatively similar to $\xi_1$] |
| **μSR:** | $\xi_{ab}$ | Fit parameter contained in the theoretical expression for $B(\mathbf{r})$ [see Eq. (3.3) and (3.4)]. |
|  | $r_0$ | Position of the maximum value of the supercurrent density $j(r) = |\nabla \times \mathbf{B}(r)|$ relative to the vortex center, measured along the straight line connecting nearest-neighbour vortices. |

## 4. Results

In this section, we discuss a selection of μSR results on different classes of type-II superconductors, with particular emphasis on the magnetic field dependence of the vortex-core size.

*4.1. V₃Si*

A good example of a strong type–II conventional *s*-wave superconductor is V$_3$Si. This A15 compound has a cubic crystal structure, a modest superconducting transition temperature $T_c = 17$ K, a large upper critical field $H_{c2} = 180$ kOe, and a GL parameter value of $\kappa \approx 27$. Small-angle neutron scattering (SANS) [26] and STS images [27] of

V$_3$Si show that for a field applied along the fourfold [001] axis the VL undergoes a continuous transformation from hexagonal below $H \approx 7.5$ kOe to square symmetry above $H \approx 40$ kOe. The nonlocal London model [20], which accounts for the fourfold symmetry of the vortices due to the anisotropy of the Fermi surface, gives a reasonable description of the VL transformation.

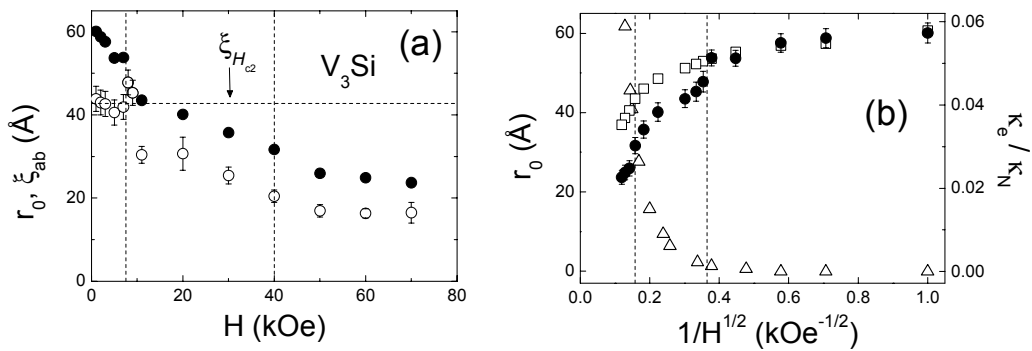

**Figure 3** (a) Magnetic field dependence of the vortex core radius $r_0$ (solid circles) and $\xi_{ab}$ (open circles) determined by μSR in V$_3$Si at $T = 3.8$ K [28]. The horizontal dashed line indicates the value of the coherence length determined from $H_{c2}$. (b) Magnetic field dependence of $r_0$ (solid circles) plotted as function of $1/H^{1/2}$, which is proportional to the intervortex spacing. The open squares represent the change in $r_0$ due to the superposition of supercurrent density profiles, calculated using the average low-field value of $\xi_{ab}$. Also shown is the electronic thermal conductivity $\kappa_e/T$ (open triangles) from Ref. [29] extrapolated to $T \to 0$ K (and normalized to the value $\kappa_N/T$ at $H_{c2}$). The dashed vertical lines indicate the field range over which the VL undergoes a continuous hexagonal-to-square transition.

Figure 3(a) shows the magnetic field dependence of $r_0$ and $\xi_{ab}$ recently determined by μSR [28]. These results were obtained from fits of the μSR time spectra to a theoretical polarization function $P(t)$ that accounts for the fourfold symmetry of the VL above 40 kOe. Consistent with the prediction of the microscopic theory, the core size $r_0$ shrinks with increasing $H$. In Fig. 3(b) a direct comparison of $r_0(H)$ to the low-temperature electronic thermal conductivity $\kappa_e(H)$ of V$_3$Si [29] is made. The data are plotted as a function of $1/H^{1/2}$, which is proportional to the intervortex spacing. The change in $r_0$ due to the vortex-lattice squeezing effect (see Section 2.2), is indicated in Fig. 3(b) by open squares. This accounts for all of the change of $r_0$ below 7.5 kOe, but there is an additional factor affecting $r_0$ at higher fields. Since the electronic thermal conductivity is sensitive only to delocalized quasiparticle states, the simultaneous increase of $\kappa_e$ and rapid decrease of $r_0$ above 7.5 kOe indicates that the vortex cores shrink further due to the intervortex transfer of quasiparticles. Thus, the distance between neighbouring fourfold-vortices in V$_3$Si at 7.5 kOe is small enough to trigger the transformation to a square VL and a simultaneous enhancement of the overlap of electronic vortex-core states.

The data for $\xi_{ab}$ in Fig. 3(a) is also very revealing. At low fields, $\xi_{ab}$ is independent of $H$ and comparable to the GL coherence length determined from $H_{c2}$, i.e. $\xi_{Hc2} =$

$(\Phi_0/2\pi H_{c2})^{1/2} \approx 42.8$ Å. On the other hand, $r_0$ decreases with increasing $H$ due to the vortex-lattice squeezing effect. With increasing $H$ above 7.5 kOe the gradual distortion of the VL is accompanied by a simultaneous reduction of $\xi_{ab}$. This contributes further to the shrinking of $r_0$. The shrinking of $\xi_{ab}$ appears to saturate above 40 kOe, where the transformation to the square VL is complete. These behaviours confirm that $\xi_{ab}$ is very sensitive to the inner structure of the vortex core defined by $\xi_1$. Consequently, $\xi_{ab} \leq \xi_{Hc2}$ over the entire field range.

Figure 4 shows the magnetic field dependence of the specific heat for the $V_3Si$ single crystal studied in Ref. [28]. Recent microscopic calculations, valid at $T = 0.1\ T_c$ [30], show that in a clean s-wave superconductor with an isotropic energy gap, the linear-$T$ coefficient $\gamma$ of the specific heat $C$ is a linear function of magnetic induction $B$ at low fields (i.e. $B/H_{c2} \leq 0.5$). This behaviour reflects the field dependence of the spatially averaged density of localized quasiparticle core states. Since the density of vortices is a linear function of $B$, so is $\gamma$. The measurements shown in Fig. 4 are at $H \leq H_{c2}/6$ and $T \approx 0.22\ T_c$. The weak nonlinearity is partly due to the diamagnetic response of the superconductor, but also the Kramer-Pesch effect at nonzero temperature. In particular, the thermally populated higher-energy core states overlap those of nearest-neighbour vortices.

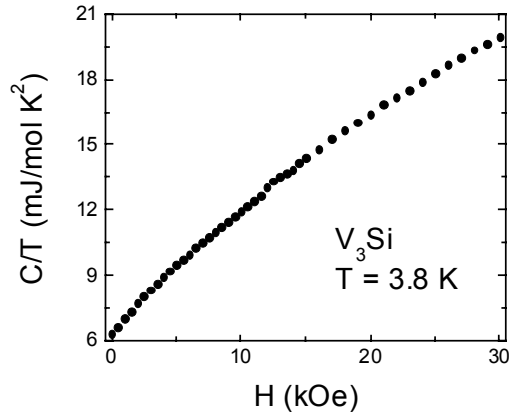

**Figure 4** The specific heat of $V_3Si$ at $T = 3.8$ K, plotted as $C(T, H)/T$ vs. $H$.

The experimental results on $V_3Si$ show that it behaves very much like the theoretical description of a conventional type-II superconductor — although we might not recognize it as such without careful consideration of the effects of field and temperature on the electronic structure of the vortices. The results presented here for $V_3Si$ serve as a good comparison for more complicated s-wave and unconventional superconductors, which are discussed next.

*4.2. NbSe$_2$*

Until recently, NbSe$_2$ was considered to be the ideal example of a simple conventional type-II superconductor. With $H$ applied parallel to the $\hat{c}$-axis direction, a hexagonal VL with long-range order is typically observed. Furthermore, NbSe$_2$ is the only superconductor in which STS has succeeded in clearly observing the localized vortex-core states [31]. However, at low fields the specific heat deviates substantially from the predicted linear-$B$ behaviour [32,33], and the core size determined by μSR shrinks rapidly with increasing applied magnetic field [34]. As shown in Fig. 5(a), the latter result is rather peculiar. At fields below $H \approx 2$ kOe the value of $\xi_{ab}$ greatly exceeds $\xi_{Hc2}$, which seems impossible.

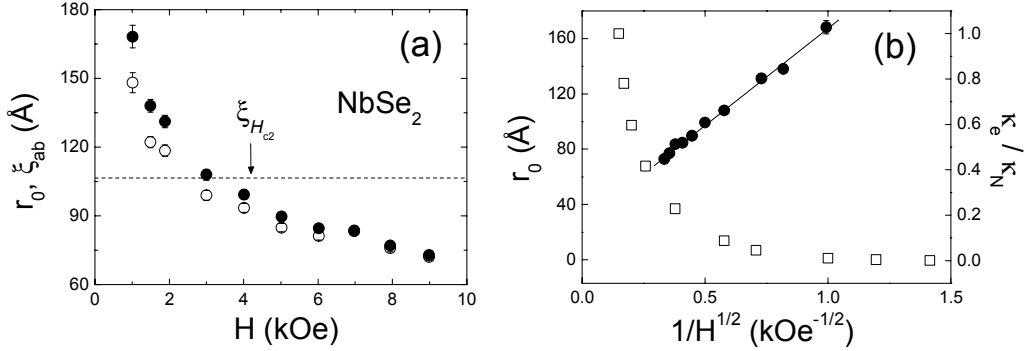

**Figure 5** (a) Magnetic field dependence of the vortex core radius $r_0$ (solid circles) and $\xi_{ab}$ (open circles) determined by μSR in NbSe$_2$ at $T = 2.3$ K [34]. The horizontal dashed line indicates the value of the coherence length determined from $H_{c2}$. (b) Magnetic field dependence of $r_0$ (solid circles) plotted as function of $1/H^{1/2}$, which is proportional to the intervortex spacing. The solid line is a guide for the eye. Also shown is the electronic thermal conductivity $\kappa_e/T$ (open squares) from Ref. [29] extrapolated to $T \rightarrow 0$ K (and normalized to the value $\kappa_N/T$ at $H_{c2}$).

Recent thermal conductivity measurements on NbSe$_2$ [29] have provided some understanding of the μSR results. Figure 5(b) shows a comparison of the field dependence of $r_0$ to the thermal conductivity data. The solid line through the data points shows that $r_0(H)$ is proportional to the intervortex spacing. Heat conduction begins to increase immediately above $H_{c1}$, indicating a high degree of delocalized quasiparticles [compare the scales in Figures 3(b) and 5(b) for $\kappa_e/\kappa_N$], while at the same time $r_0$ (and $\xi_{ab}$) decrease. In Ref. [29], these behaviours were attributed to superconductivity on two different sheets of the Fermi surface characterized by a larger and smaller energy gap, denoted here as $\Delta_L$ and $\Delta_S$, respectively. The ratio $\Delta_L/\Delta_S$ is nearly that of the low and high-field values of $r_0$, indicating that the vortex cores are governed by length scales $\xi_S \sim 1/\Delta_S$ (at low field) and $\xi_L \sim 1/\Delta_L$ (at high field). This situation is similar to the two-gap superconductor MgB$_2$, where STS measurements on the π band at $H \sim 0.02\,H_{c2}$ show that the vortex cores are much larger than the size estimated from $H_{c2}$ [35]. Nakai, Ichioka and Machida [36] have developed a two-band model for MgB$_2$, which may be used here to understand the behaviour of $r_0$ in NbSe$_2$. In this picture, quasiparticle states at low fields are highly confined to the L-band vortex, but loosely bound to the S-band vortex. In other words, the low-energy cores states in the S-band extend well outside the vortex core. This results in a significant overlap of the core states of nearest-neighbour vortices at low fields. With increasing $H$, the enhanced intervortex transfer of quasiparticles in the S-band results in a rapid shrinking of $r_0$. At higher field, $r_0$ reflects the core size of the L-band vortex. Thus, the occurrence of two energy gaps is a natural explanation for the stronger than expected field dependence of the core size in $NbSe_2$.

The modification of the vortex core electronic spectrum due to vortex-vortex interactions should have a profound effect on the strength of the observed Kramer-Pesch effect. Figure 6 shows the temperature dependence of $r_0$ for $NbSe_2$ at two values of $H$ [37,38]. The smaller vortex-core size at 5 kOe can be understood as arising from the vortex-lattice squeezing effect and the increased overlap of the quasiparticle core states of neighbouring vortices. The thermal depopulation of bound core state energy levels is predicted to saturate at a temperature where the thermal smearing is narrower than the energy-level spacing [8-11]. This is clearly observed in the µSR data at $H$ = 5 kOe. On the other hand, the larger vortex cores at 1.9 kOe contain a greater number of discrete energy levels, so we expect the shrinkage of the vortex cores to saturate at a lower temperature. While the measurements at 1.9 kOe do not go to low enough temperature to observe complete saturation of $r_0(T)$, the larger slope of the data compared to 5 kOe is consistent with a reduced saturation temperature. The larger slope at 1.9 kOe also indicates that the strength of the Kramer-Pesch effect increases for larger vortex separation. This tendency is supported by microscopic calculations for *isolated* vortices [8-11], where the temperature dependence of the vortex-core size is determined to be stronger than the µSR results shown in Figure 6.

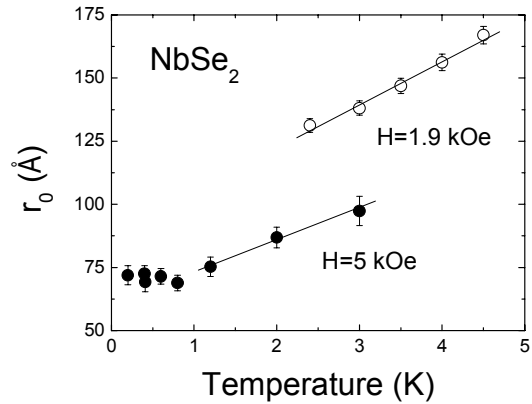

**Figure 6** Temperature dependence of the of the vortex-core size $r_0$ in $NbSe_2$ determined by µSR. The open circles are data taken at $H$ =1.9 kOe [37] and the solid circles are data taken at $H$ =5 kOe [38]. The solid lines provide a guide to the eye.

*4.3. $CeRu_2$*

Experimental findings in the superconducting state of $CeRu_2$ are for the most part consistent with *s*-wave BCS-like behaviour. SANS [39] and STS [40] studies indicate that the VL in $CeRu_2$ has hexagonal symmetry, at least up to $H$ = 20 kOe. A µSR study by Kadono *et al.* showed that the vortex cores shrink in size with increasing magnetic field [41]. As in $V_3Si$, the value of $\xi_{ab}$ decreases below $\xi_{Hc2}$ with increasing magnetic field — indicating a change in the spatial derivative of $B(r)$ and hence the slope of $\Delta(r)$ at the center of the vortex cores. Thus, the field dependence of $r_0$ can be attributed to both the vortex-lattice squeezing effect and the intervortex transfer of quasiparticles. However, it must be noted that in contrast to theoretical expectations for a conventional *s*-wave superconductor [30], the specific heat for $CeRu_2$ exhibits a sizeable nonlinear field dependence at low $H$ and $T \approx 0.1\ T_c$ [42]. Assuming $CeRu_2$ is indeed an *s*-wave superconductor, this raises the possibility that the superconducting properties of $CeRu_2$ at low $H$ are governed by a smaller energy gap on a different part of the Fermi surface — as is the case for $NbSe_2$ and $MgB_2$. However, this appears not to be the case. As shown in Fig. 7, the value of $\xi_{ab}$ determined by μSR at low fields is comparable to $\xi_{Hc2}$. This behaviour is similar to $V_3Si$ (see Fig. 3), but is very different from the low-field behaviour of $\xi_{ab}(H)$ in $NbSe_2$, where $\xi_{ab} > \xi_{Hc2}$ (see Fig. 5). In Ref. [41] it was shown that by assuming the conventional picture of localized quasiparticle core states, the field dependence of the linear-*T* coefficient γ of the specific heat at $T \approx 0.33\ T_c$ can be explained by the shrinkage of the vortex cores with increasing field. Thus, the μSR measurements are consistent with $CeRu_2$ being a single-band conventional *s*-wave superconductor.

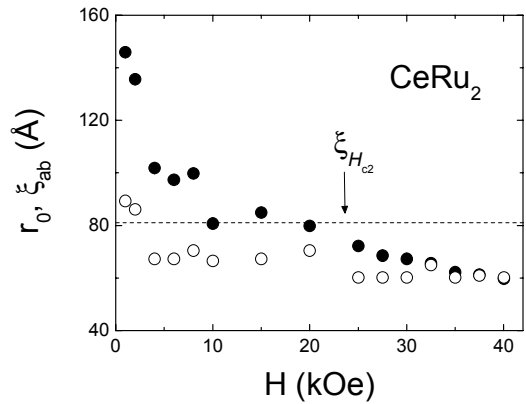

**Figure 7** Magnetic field dependence of the vortex core radius $r_0$ (solid circles) and $\xi_{ab}$ (open circles) determined by μSR in $CeRu_2$ at $T = 2.0$ K [41]. The horizontal dashed line indicates the value of the coherence length determined from $H_{c2}$.

*4.4. $LuNi_2B_2C$ and $YNi_2B_2C$*

The energy gap in the non-magnetic tetragonal borocarbide superconductors $LuNi_2B_2C$ and $YNi_2B_2C$ is known to be highly anisotropic. Recent thermal conductivity measurements on $YNi_2B_2C$ suggest that the gap function has point-like minima [43]. The near $H^{1/2}$ dependence of the electronic specific heat observed in $LuNi_2B_2C$ [44,45] and $YNi_2B_2C$ [33] supports this picture. In particular, there is a significant increase in quasiparticle excitations outside the vortex cores due to a Doppler shift of the quasiparticle energy spectrum [46]. With increasing magnetic field applied along the fourfold tetragonal axis, the VL in these borocarbide superconductors undergoes a hexagonal-to-square symmetry transition, followed by a transition back to hexagonal [47]. The fields for these transitions reported in the literature are somewhat sample dependent. The low field hexagonal-to-square VL transition is in reasonable agreement with the nonlocal London model [48].

The magnetic field dependence of the vortex-core size $r_0$ has been determined by µSR in $LuNi_2B_2C$ [49] and $YNi_2B_2C$ [50]. The data analysis in both studies assumed that $B(\mathbf{r})$ is described by the nonlocal London model. For both samples, $r_0$ is significantly larger than $\xi_{ab}$. As discussed in Section 4.1, we expect $r_0$ to exceed $\xi_{ab}$ due to the effects of overlapping $j(r)$ profiles of opposite sign. The difference is particularly large in this case, because µSR measurements on these borocarbide superconductors also show a strong field dependence for $\lambda_{ab}$ (see article by Kadono in this issue). The field dependence of $\lambda_{ab}$ further modifies the spatial overlap of the $j(r)$ profiles from nearest-neighbour vortices.

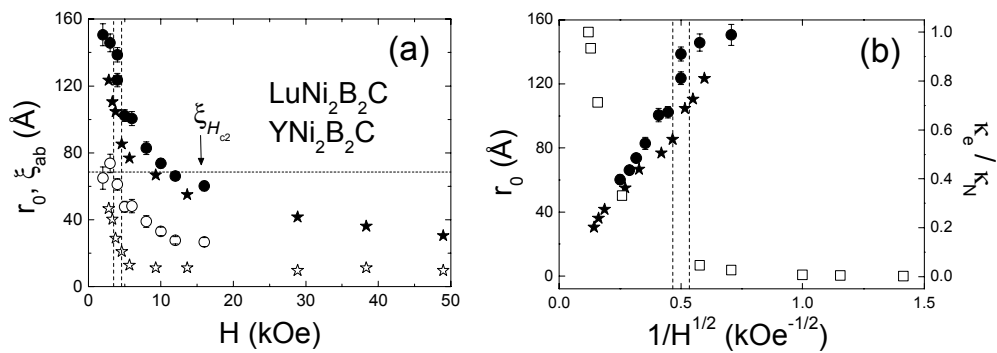

**Figure 8** (a) Magnetic field dependence of the vortex core radius $r_0$ determined by µSR in $LuNi_2B_2C$ at $T = 2.5$ K [49] (solid circles) and in $YNi_2B_2C$ at $T = 3.0$ K [50] (solid stars). Also shown is $\xi_{ab}$ for $LuNi_2B_2C$ (open circles) and $YNi_2B_2C$ (open stars). The horizontal dashed line indicates the value of the coherence length determined from $H_{c2}$ and the vertical dashed lines approximately indicate the field range over which the VL undergoes a continuous hexagonal-square transition. (b) Magnetic field dependence of $r_0$ (solid circles and stars) plotted as function of $1/H^{1/2}$. Also shown is the electronic thermal conductivity $\kappa_e/T$ (open squares) for $LuNi_2B_2C$ [51] extrapolated to $T \to 0$ K (and normalized to the value $\kappa_N/T$ at $H_{c2}$).

The rapid drop of $\xi_{ab}$ with increasing $H$ suggests that the inner structure of the vortex core changes rapidly at low fields due to the delocalization of bound core states. This is consistent with a highly anisotropic gap, such that $\xi_{ab}$ is inversely proportional to the gap minima at low fields. Similar to $V_3Si$ and $CeRu_2$, the value of $\xi_{ab}$ does not exceed the value of the coherence length calculated from $H_{c2}$ at any field. However, it is conceivable that $\xi_{ab}$ exceeds $\xi_{Hc2}$ at fields lower than the data shown in Fig. 8(a). The reduction of $\xi_{ab}$ with increasing field saturates above the hexagonal-to-square VL transition. This suggests the core states are *completely delocalized* at these higher fields. Consequently, the continued shrinking of $r_0$ with increasing $H$ is due entirely to the vortex-lattice squeezing effect. The strong reduction of $r_0$ and corresponding increase of the electronic thermal conductivity [51] above the VL transition [see Fig. 8(b)] are thus attributable to the excitation of extended quasiparticle states outside of the vortex cores.

*4.5. $YBa_2Cu_3O_{7-\delta}$*

In high-$T_c$ superconductors, the structure of the VL is expected to be strongly influenced by the *d*-wave order parameter (see Ref. [18] and references therein). Theoretically, the VL in a *d*-wave superconductor is characterized by fourfold symmetry and quasiparticle core states that are extended along the nodal directions of the order parameter. Specific heat measurements seem consistent with this picture [52-54] and a fourfold symmetric VL has been observed in $La_{1.83}Sr_{0.17}CuO_{4+\delta}$ (LSCO) [55] and $YBa_2Cu_3O_{7-\delta}$ (YBCO) [56,57] by SANS. However, in LSCO the orientation of the VL is 45° with respect to the Cu-O bonds, which does not agree with the theoretical predictions for a *d*-wave VL. This at least raises the possibility that the fourfold symmetry of the VL in LSCO arises from strong coupling to the underlying Fermi surface, rather than the *d*-wave symmetry of the order parameter. In untwinned $YBa_2Cu_3O_{7-\delta}$, an hexagonal VL distorted by strong *a-b* anisotropy due to the CuO chains is observed by SANS at fields $H \leq 40$ kOe directed along the $\hat{c}$-axis [53,54]. The VL gradually distorts with increasing field above 40 kOe, transforming into a square lattice near 110 kOe.

Figure 9(a) shows the field dependence of the vortex-core size extrapolated to $T = 0$ K in $YBa_2Cu_3O_{6.95}$ [25]. These results came from an analysis of the μSR time spectra that assumed the VL has hexagonal symmetry over the entire field range. In light of the recent findings by SANS, there is some additional uncertainty in the absolute values of $\xi_{ab}$ and $r_0$ at fields above 40 kOe. With increasing applied magnetic field, $r_0$ decreases rapidly and saturates over the field range where the gradual transition to a square VL takes place. The rapid decrease of $r_0$ is partly due to the strong field dependence of $\lambda_{ab}$, which is simultaneously determined in the analysis of the μSR data [25]. Increasing $\lambda_{ab}$ results in a greater overlap of the $j(r)$ profiles of neighbouring vortices in Fig. 2, and hence a reduction of $r_0$. Consequently, there is a definite correlation between the field dependence of $r_0$ and $\lambda_{ab}$. In particular, from the data in Ref. [25], $\lambda_{ab} \propto H^{1/2}$ and $r_0 \propto 1/H^{1/2}$ [see Fig. 9(b)], so that $\lambda_{ab}/r_0 \propto H$. Since $\xi_{ab}$ also decreases with increasing $H$, this also contributes to the field dependence of $r_0$.

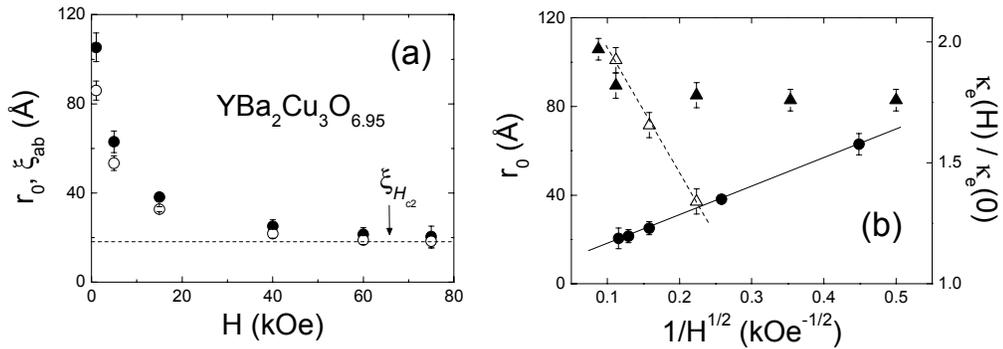

**Figure 9** (a) Magnetic field dependence of the vortex core radius $r_0$ (solid circles) and $\xi_{ab}$ (open circles) determined by µSR in $YBa_2Cu_3O_{6.95}$ extrapolated to $T \to 0$ K [25]. The horizontal dashed line indicates the value of the coherence length determined assuming $H_{c2} = 1000$ kOe. (b) Magnetic field dependence of $r_0$ (solid circles) plotted as function of $1/H^{1/2}$. Also shown is the normalized electronic thermal conductivity $\kappa_e(H)/\kappa_e(0)$ extrapolated to $T \to 0$ K, from Ref. [58] (open triangles) and Ref. [59] (solid triangles). The solid and dashed lines are guides for the eye.

Figure 9(b) shows a comparison of $r_0$ to the electronic thermal conductivity measured in $YBa_2Cu_3O_{6.9}$ [58], and more recently in ultra-clean $YBa_2Cu_3O_7$ [59]. A comparison with the data of Ref. [58] would seem to imply that delocalization of bound quasiparticle core states contributes to the field dependence of $r_0$. However, the dominant contribution to $\kappa_e$ likely comes from a field-induced Doppler shift of the quasiparticle spectrum outside the vortex cores [46]. Furthermore, the field dependence of $\kappa_e$ in the ultra-clean sample is essentially nonexistent — likely due to an exact cancellation of the contributions to the thermal conductivity from the Doppler shift and scattering of quasiparticles off the vortices [59]. Thus, there is no correlation in this case between thermal conductivity and µSR measurements of the vortex-core size.

A puzzling feature of the data in Fig. 9(a) is the large size of the vortex cores at low fields. On the other hand, above 40 kOe the value of $\xi_{ab}$ is less than 20 Å, which is certainly consistent with the large value of $H_{c2}$ in $YBa_2Cu_3O_{6.95}$. It is also interesting to note that the VL begins its gradual hexagonal-to-square transition at 40 kOe [57]. Recently it has been shown that the same data analysis that yielded the low-field value of the core size in Fig. 9(a) also yielded accurate values of the *absolute value* of $\lambda_{ab}$ — in excellent agreement with Gd-ESR spectroscopy measurements [60]. In addition, the temperature dependence of $\lambda_{ab}$ at *low fields* has been shown [25] to agree with microwave cavity perturbation measurements in the Meissner phase [61]. Thus, the large vortex-core size observed at low fields in $YBa_2Cu_3O_{6.95}$ is of intrinsic origin and not an artifact of the modeling procedure.

Based on the survey of superconductors in the previous sections, the large value of $\xi_{ab}$ at low fields most likely originates from sensitivity to a smaller energy gap. One possibility is that because breaking of the superconducting pairs at low fields is dominated by quasiparticles excitations near the nodal regions of the *d*-wave energy gap, $\xi_{ab}$ is inversely proportional to the average gap in the vicinity of the nodes. With increasing $H$ there is an increasing contribution from antinodal quasiparticles, so that at higher fields $\xi_{ab}$ reflects the larger gap on these regions of the Fermi surface. However, µSR measurements on $La_{1.85}Sr_{0.15}CuO_4$ single crystals [62] seem to rule out this explanation. The analysis of the data taken at $H = 2$ kOe and $T = 1.2$ K in Ref. [62] yielded $\xi_{ab} \approx 51$ Å, which is comparable to $\xi_{Hc2}$ for LSCO.

Since the large vortex-core size is not a universal property of the cuprates, there must be another source unique to YBCO. It seems likely then that the one-dimensional CuO chains are responsible. Atkinson and Carbotte [63], and Xiang and Wheatley [64] have developed proximity models for YBCO in which a small energy gap is induced on the chains by the superconducting $CuO_2$ planes. Xiang and Wheatley [64] have also considered a model whereby the chains and planes are coupled through the tunneling of electron pairs. Alternatively, Whelan and Carbotte [65] have constructed a model whereby the effect of the CuO chains is to produce a modified gap function that applies to both the chains and planes. They predict that this introduces a low-energy scale into the quasiparticle excitation spectrum characterizing the superconducting energy gap at fields below ~ 45 kOe. The authors of Ref. [65] suggest that the low-energy scale corresponds to the anomalous feature observed in the local density of states (LDOS) by STS at 5.5 meV [66]. Taking the upper scale to be the *d*-wave gap maximum ~ 20 meV, $\xi_{ab}$ at the lowest field should be roughly a factor of 4 times larger than at high fields. This is precisely what the data in Fig. 9(a) shows. Furthermore, the saturation in the shrinking of $\xi_{ab}$ is near the predicted field at which the magnetic energy sweeps through the van Hove singularity in the LDOS at 5.5 meV. Thus, we can understand the large vortex-core size in YBCO at low fields as being due to the CuO chains. It must be said that the physical picture suggested in Ref. [65] is probably not correct, and a gap on the CuO chains is more likely to originate from the proximity effect. Nevertheless, one expects the LDOS derived from proximity models to resemble that of Ref. [65], so that the experimental signatures are similar. Finally, like the borocarbides, the saturation of $\xi_{ab}$ observed at high fields by μSR suggests that the quasiparticle states are not bound to the vortex cores. This is consistent with the weak Kramer-Pesch effect also observed by μSR [37].

## 8. Summary and Conclusions

From a close inspection of μSR measurements and a comparison with findings from other experimental techniques, it has been established that the size of the vortex cores in a superconductor is affected by (*i*) the vortex-lattice squeezing effect, (*ii*) the intervortex transfer of quasiparticles, and (*iii*) changes to the nature of the superconducting energy gap [which affects (*i*) and (*ii*)]. While the calculated position of the peak in the supercurrent density $r_0$ in the μSR experiments is influenced by all three factors, the fitting parameter $\xi_{ab}$ appears to be sensitive only to the latter two. A striking conclusion of this survey is that the large vortex-core size in YBCO at low fields originates from a low-energy scale imposed by the CuO chains.

## Acknowledgements

I am most grateful to D. Broun, Z.X. Shen, K. Machida, J. Carbotte and F.D. Callaghan for helpful discussions. This work was supported by the Natural Sciences and Engineering Research Council (NSERC) of Canada and the Canadian Institute for Advanced Research (CIAR).